\documentclass[fleqn,usenatbib,letters]{mnras}

\usepackage[T1]{fontenc}
\usepackage{ae,aecompl}
\usepackage{xcolor}

\usepackage{graphicx}	
\usepackage{amsmath}	
\usepackage{amssymb}	

\title[Radio jets in NGC 5322]{LOFAR discovery of rare large FR-I jets in low-luminosity radio galaxy NGC 5322}

\author[A. Omar]{
Amitesh Omar\thanks{E-mail: aomar@aries.res.in (AO)}
\\
Aryabhatta Research Institute of observational-sciences, Manora Peak, Nainital, 263001, India\\
}

\begin{document}
\maketitle

\begin{abstract}

The discovery of faint FR~I radio jets in the nearby elliptical galaxy NGC 5322 is reported here using 144 MHz LOFAR image. The jets have an angular extent of $\sim40$ arcmin or a projected physical extent of $\sim350$ kpc. The faint jets remain well collimated and disappear in the intergalactic medium, without any visible hotspot or radio lobes. The jets detected up to $\sim20$ kpc extent at higher frequencies are relatively bright within the optical extent of the galaxy but become faint abruptly outside, where detection is made only in the LOFAR image. The total radio luminosity of the galaxy at 144 MHz is estimated to be $3.7(\pm0.4)\times10^{22}$ W Hz$^{-1}$. The 144 MHz radio luminosity of the faint jets outside the optical extent is estimated to be $7.1(\pm2.0)\times10^{21}$ W Hz$^{-1}$. The extent of the jets for its radio luminosity is abnormally large when compared to the general population of radio galaxies. It makes NGC 5322 a member of a rare population of radio galaxies, previously not detected in other radio surveys. A combined effect of stellar core depletion and low-density environment around the jets, inferred from previous studies in other wave-bands, resulting into weak entrainment of surrounding material to the jets could be responsible for its large size despite a low radio luminosity.    
\end{abstract}

\begin{keywords}
galaxies: active -- galaxies: jets -- radio continuum: galaxies -- (galaxies:) intergalactic medium -- galaxies: nuclei
\end{keywords}



\section{Introduction}

NGC~5322 is an optically bright ($m_{\mathrm{r}} \sim 10.3$ mag; $M_{\mathrm{r}} \sim -22.2$ mag) nearby (distance $\sim31$ Mpc; H$_{0}$ = 70~km~s$^{-1}$~Mpc$^{-1}$; scale $1'\sim8.7$ kpc) elliptical (E$3-4$; \citealt{RC3}) radio galaxy of Low Ionization Nuclear Emission Region (LINER) type \citep{Baldi}. It was detected in early radio interferometric surveys using the Westerbork Synthesis Radio Telescope (WSRT) and the Very Large Array (VLA) at frequencies of 1.4 GHz and 5.0 GHz \citep{Sramek, Hummel80, Hummel84}. These observations revealed two {\it classical} bipolar radio jets on kpc-scales and a core. The radio jets of extent $\sim2.3\arcmin$ ($\sim20$ kpc) at a P.A. of nearly $-7^{\circ}$ in the 1.4 GHz WSRT image at a resolution $15\arcsec\times13\arcsec$ and detection sensitivity of 0.5 mJy beam$^{-1}$ were found to be completely embedded within the optical extent of the galaxy \citep{Feretti84}. \citet{Nagar} identified core-jet morphology in NGC 5322 using the VLBA (Very Large Baseline Array) observations. \citet{Dullo} presented a high angular resolution study of NGC 5322 at 1.5 GHz using the e-MERLIN array and optical observations using the {\it HST}. The {\it HST} image revealed an edge-on nuclear dust disk, aligned with the major axis of the galaxy and nearly perpendicular to the pc-scale radio jets detected in the e-MERLIN image.

The 1.4 GHz flux of NGC 5322 in the NVSS (NRAO VLA Sky Survey; \citealt{NVSS}) is $78\pm3$ mJy, which is in good agreement with the earlier 1.4 GHz WSRT estimate of $84\pm4$ mJy by \citet{Feretti84}. The 1.4 GHz FIRST (Faint Images of the Radio Sky at Twenty-Centimeters; \citealt{FIRST}) survey image shows a core-jet morphology at 5\arcsec resolution. The galaxy is also detected at 325 MHz in the Westerbork Northern Sky Survey (WENSS; \citealt{WENSS}), 150 MHz in the TIFR GMRT Sky Survey (TGSS;  \citealt{TGSS}), and at 74 MHz in the VLA Low-Frequency Sky Survey Redux (VLSSr; \citealt{VLSSr}) images. The source is slightly resolved in the NVSS and TGSS images and un-resolved in the VLSSr image. The total radio flux densities of NGC 5322 measured in the images from different radio surveys are provided in Table~\ref{tb:flux}. 

The radio emission from NGC 5322 was examined in the recently released LOFAR (Low Frequency ARray; \citealt{LOFAR}) Two-metre Sky Survey (LoTSS) DR2 images at 144 MHz \citep{LoTSS}. The LoTSS is being carried out using the LOFAR, The Netherlands, using wide-bandwidth ($120 - 168$ MHz) high-band antennas. The calibrated and de-convolved images are provided in the LoTSS image archives. These images have high fidelity and capable of detecting extended emission. The flux density scale in the LoTSS-DR2 survey is expected to be accurate within $\pm10$ per-cent. NGC 5322 is detected in the LoTSS data mosaic P207+60 with an angular offset of 0.27 degree from the mosaic center. 

The detection of faint radio jets in NGC 5322 on hundred kpc-scales in the LoTSS image is reported in this paper. The astrophysical scenarios for its origin are discussed here. We used cosmological parameters H$_{\mathrm 0}$ = 70~km~s$^{-1}$~Mpc$^{-1}$, $\Omega_{\mathrm M}=0.3$ and a flat Universe to derive distances. 

\section{LOFAR radio morphology}

\begin{table}
\begin{center}
\caption{Radio flux of NGC 5322}
\label{tb:flux}
\begin{tabular}{lcccr}
\textbf{Frequency} & \textbf{Total Flux} & \textbf{LAS$^a$} & \textbf{rms} & \textbf{Survey}\\
      MHz & mJy & arcsec & mJy beam$^{-1}$ & \\
      \hline
      74 & $432\pm51$ & 1100 & 65 & VLSSr \\
      144 & $266\pm27^{c}$ & 2578 & 0.07 &  LoTSS \\
          & $36\pm9^{N}$ & & & \\
          & $25\pm8^{S}$ & & & \\
      150 & $217\pm22$ & 4080 & 2.8 & TGSS \\
      327 & $144\pm6$ & 1375 & 3.8 & WENSS \\
      1400 & $78\pm3$ & 970 & 0.5 & NVSS\\
 \hline
       \end{tabular}
  \end{center}
  \footnotesize{Note: $a$ Largest Angular Scales (LAS) that may be reliably detected in interferometric surveys are taken from \citep{Savini} for all the surveys but TGSS, which is taken from \citep{TGSS}. $c$ indicates flux density for bright jet portion within $\pm1.1'$ from the center measured in this work; $N$ and $S$ indicates flux densities for Northern and Southern jets respectively.}
\end{table}

\begin{figure}
\includegraphics[width=\columnwidth]{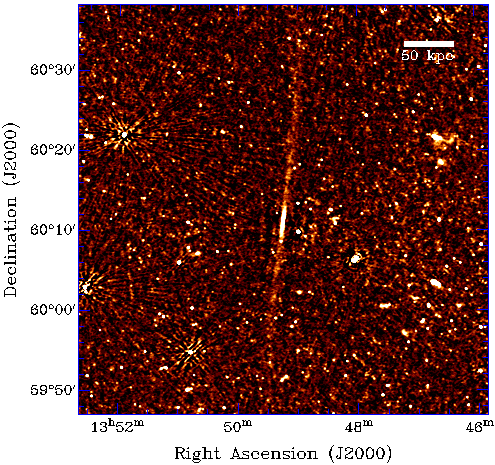}
\caption{144 MHz LoTSS image of NGC 5322. The image is taken from high-resolution mosaic and later smoothed using a 3 pixel width Gaussian. The rms in the raw image is $70\mu$Jy beam$^{-1}$ at an angular resolution of $5"$.}
\label{fig:lofar}
\end{figure}

\begin{figure}
\includegraphics[width=\columnwidth]{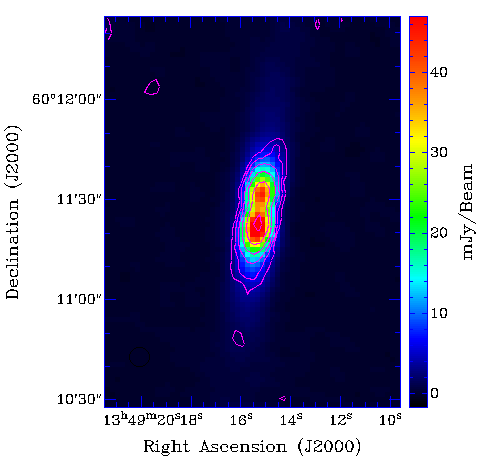}
\caption{Contours from 1.4 GHz FIRST image overlaid upon 144 MHz LoTSS image of NGC 5322. The contour levels are 0.45 mJy beam$^{-1} \times (1,2,4,8,16,32)$.}
\label{fig:first}
\end{figure}

\begin{figure}
\includegraphics[width=\columnwidth]{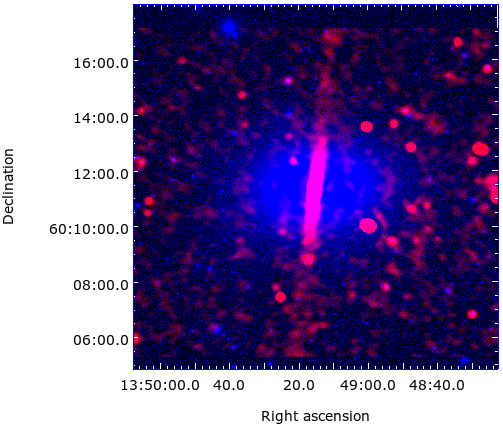}
\caption{Color overlays of SDSS i-band image (blue) and LoTSS image (red) highlighting central bright emission region within the optical extent of NGC 5322.}
\label{fig:sdss}
\end{figure}

Fig.~\ref{fig:lofar} shows the 144 MHz high-resolution ($6\arcsec\times6\arcsec$) LoTSS image of NGC 5322. The image is contrast-adjusted on non-linear scales to reveal faint diffuse emission in presence of bright radio sources. This image is smoothed to $3\times3$ pixels using a Gaussian kernel to further enhance diffuse radio emission. Two {\it classical} jet-like features emanating from the galaxy are detected. The jets are bright within $\pm1$ arcmin from the center and afterwards become faint abruptly. The faint jet-like features on tens of arcmin-scale cannot be de-convolution errors as these errors have a different pattern, which can be seen around bright point sources in the image. The jets are nearly North-South oriented and detected between the declinations $59^{\mathrm d}~52^{\mathrm m}$ and $60^{\mathrm d}~32^{\mathrm m}$. The jet emission appears fragmented on smaller scales, however, due to poor signal to noise ratio, details cannot be quantified. The angular extent of the jets ($\sim40\arcmin$) is close to the Largest Angular Scale (LAS) that the LOFAR can possibly sample (see e.g., \citealt{Savini}). Therefore, the true extents of the jets may be larger than that detected in the LoTSS image. The typical LAS values for the radio surveys are provided in Tab.~\ref{tb:flux}. 

The FIRST image being at a comparable angular resolution to that in the LoTSS image, an overlay is shown in Fig.~\ref{fig:first}. The bright regions of NGC 5322 in the two images have very similar radio morphologies with identical locations for the brightest emission region. The jets detected in the 1.4 GHz WSRT image \citep{Feretti84}, within $\pm1.1$ arcmin from the center are detected as bright jets in the 144 MHz LoTSS image also, over roughly the same extent. The faint jets on 100-kpc scale detected in the LoTSS image are not detected in the previously available images. The total flux at 144 MHz is measured for the central bright portion of the jets over $\pm1.1'$ from the center so that it can be compared with the measurements from the past surveys, which have not detected outer faint jets. The average spectral index ($\alpha$; where $S\propto\nu^{\alpha}$) of the bright jet (within $\pm1.1$ arcmin) is estimated to be $-0.67\pm0.14$ between 1.4 GHz and 144 MHz, using the flux values provided in Tab.~\ref{tb:flux}. The radio emission is therefore of synchrotron origin, typical for radio galaxies.  It was assumed here that all other surveys detected the bright jets although due to poor angular resolution it is not verifiable. We will refer the faint jets hereafter as those detected in the LoTSS image only. The quoted errors in the flux density values at 144 MHz are estimated as combination of rms in the summed region and a 10 per-cent calibration error. The flux estimated from the 150 MHz TGSS image is marginally different compared to that from the 144 MHz LoTSS image, although the values agree within the flux calibration errors in the two surveys.  

The average radio surface brightness of the faint jets is measured as $\sim76~\mu$Jy arcsec$^{-2}$, which is nearly at a level of $3\sigma$ detection limit of the LoTSS images. The faint jets are not detected in the TGSS due to much higher rms in the TGSS despite having LAS sensitivity. The faint jets are not detected in other surveys due to limitations arising from both rms and LAS sensitivities and the jets getting fainter at higher frequencies. The surface brightness sensitivity limit in the NVSS is estimated as $\sim1.4~\mu$Jy arcsec$^{-2}$ for a flux sensitivity $3\sigma\sim1.4$ mJy beam$^{-1}$, beam size $45''\times45''$, and beam area dilution by a factor 2 as the jet width is not more than $20''$ in most regions. Therefore, the non-detection in the NVSS image implies that the faint jets are steeper than $\alpha\sim-1.7$. An attempt was also made to obtain spectral index of the faint jets using three in-band LoTSS images having smaller bandwidths, however due to low fidelity and poor signal to noise ratio in lower bandwidth images, features appeared very noisy hence not presented here. It will be worth to obtain a deeper image of the faint jets using LOFAR or GMRT at meter-wave bands and determine its spectral index.

A color overlay of the LoTSS image and the optical i-band image from the Sloan Digital Sky Survey (SDSS) is shown Fig.~\ref{fig:sdss}. The jets seem to abruptly become faint after leaving the bright optical regions of the galaxy. The faint jets remain well collimated over a long range and appear to diffuse out in the outer regions before disappearing (see Fig.~\ref{fig:lofar}). The projected physical length of the jets is estimated to be $\sim350$ kpc. The Northern faint jet (flux=$36\pm9$ mJy) appears relatively brighter than the Southern faint jet (flux=$25\pm8$ mJy). If this difference is due to Doppler boosting, the faint jets must be moving at relativistic speeds. 
 
\section{Discussions}

\begin{figure}
\includegraphics[width=\columnwidth]{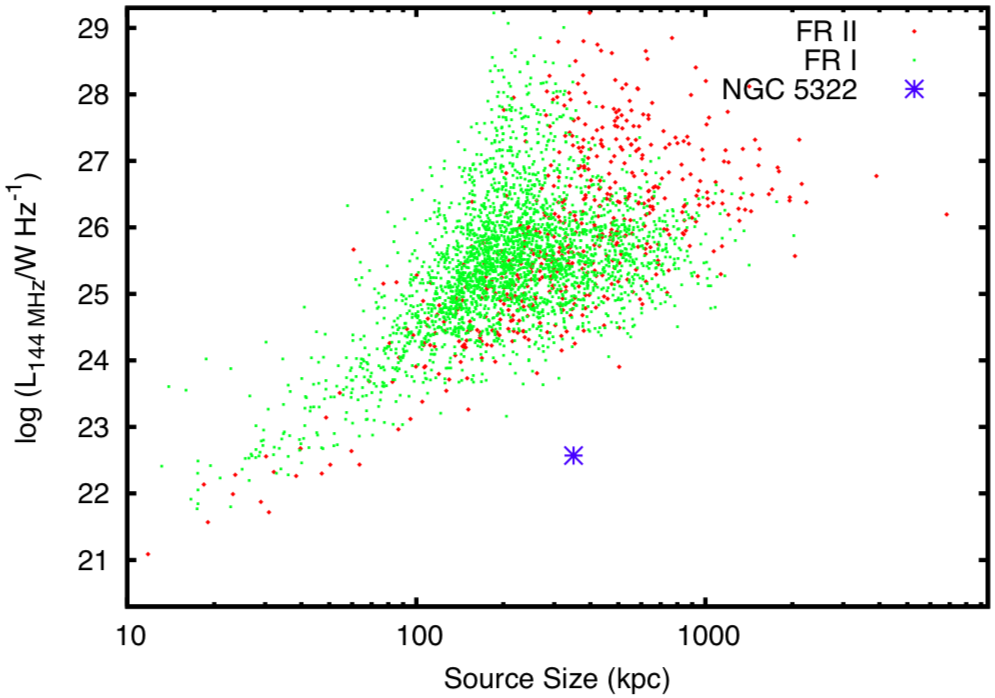}
\caption{Location of NGC 5322 in the luminosity-size plot for FR~I and FR~II galaxies detected in the LoTSS DR1 and first made available in \citet{Mingo}. The origins of redshifts are spectroscopic (wherever available) or else photometric. The luminosities are K-corrected with a spectral index of $-1$. The symbols are explained in the upper right corner.}
\label{fig:L-D}
\end{figure}

The 144 MHz radio morphology of NGC 5322 appears that of the classical FR~I type (see \citealt{FR}) as the jets are edge-darkened. The 1.4 GHz radio luminosity using the NVSS flux is estimated as $\sim1\times10^{22}$ W Hz$^{-1}$ that makes NGC 5322 as a low luminosity radio galaxy. The radio luminosity of the large-scale faint radio jets outside the optical extent, detected only at 144 MHz, is estimated to be $7.1(\pm2.0)\times10^{21}$ W Hz$^{-1}$, and that of all radio emission at 144 MHz, associated with NGC 5322 is estimated to be $3.7(\pm0.4)\times10^{22}$ W Hz$^{-1}$. The projected end-to-end extent of the radio jets is $\sim350$ kpc. For its radio luminosity, this source size is exceptionally large as can be seen from Fig.~\ref{fig:L-D}, where luminosities and sizes are plotted for FR~I and FR~II galaxies up to $z\sim2$, detected in LoTSS DR1, using  data provided in \citet{Mingo} and \citet{Will}. It can be seen from Fig.~\ref{fig:L-D} that for sources with their sizes similar to that of NGC 5322, typical radio luminosity of  galaxies is more than one order of magnitude higher than that of NGC 5322. Same inference could also be drawn from an earlier luminosity-size plot made by \citet{An} at 1.4 GHz. Therefore, NGC 5322 is a rare radio galaxy, detected for the first time to the best of our knowledge having oversized radio jets with at least one order of magnitude lower radio luminosity than the normal. The 100-kpc scale radio jets in NGC 5322 are detected for the first time, thanks to the unprecedented sensitivity of the LoTSS survey at low radio frequencies where jets become bright.

Other important features of the jets in NGC 5322 are its abrupt dimming outside the optical extent of the galaxy and its collimation on 100-kpc scales. The jets in radio galaxies almost always propagate at relativistic speeds. The emissivity of jets are affected by various particle energy loss processes such as synchrotron emission, adiabatic expansion, and inverse-Compton scattering by cosmic microwave background photons (see e.g., \citealt{An}). The efficiency of different energy loss processes can vary across the length of jet, e.g., synchrotron losses will dominate in highly relativistic jets in high magnetic fields while inverse-Compton loss will dominate in outer ends of jets and lobes with low magnetic field strengths. The evolutions of jets and its magnetic field are complex and can depend on various parameters such as central black hole (BH) properties (e.g., mass, spin, accretion), initial jet power, physical conditions of interstellar medium (ISM) and inter galactic medium (IGM) surrounding the jets (see review by \citealt{Hard} and \citealt{Saikia}). 

The predicted spectral index of the faint jets is highly steep ($\alpha\ge-1.7$) as shown in the previous section. A steep spectrum may be interpreted as old plasma hence may support a scenario where the faint jets are relic jets in which high energy electrons emitting beyond the synchrotron break frequency have lost energy and only long-lived (age $\sim100$ Myr) low energy electrons are emitting at very low frequencies. However, it is then surprising that (i) such old relic jets still remain well collimated and (ii) moving at bulk relativistic speeds if one considers the observed difference in the radio flux of bipolar jets due to the Doppler boosting effect. The faint jets detected at 144 MHz appear continuous on large scale. The faint jets also connect seamlessly with the relatively brighter jets that are also detected at GHz frequencies. This connection may be interpreted as continuous supply of energy to the jets and hence the faint jets may not be old relics from a previous episode of jet activity. In a case where we consider bright jets as due to a more recent activity, then the new brighter jets are required to be almost exactly aligned in the same direction as that of the old jets. Normally, jet/BH precession and galaxy motions over a long time are expected to change the direction of jets significantly. With the limited data in hand, it is not straightforward to understand brightness evolution of radio jets in NGC 5322. A detailed study using multi frequency data with LOFAR and GMRT will be very valuable to study jet physics in this object.

The jets in FR~I radio galaxies are modeled as turbulent decelerating flows. In a classical picture, low power FR~I jets slow down, become sub-relativistic, de-collimate and disperse into a wider jet or diffuse radio lobes beyond a few tens of kpc distance from the parent galaxy (see e.g., \citealt{Laing}). On the other hand, more powerful FR~II jets remain well collimated and are able to create bright hotspots and radio lobes up to several hundreds of kpc distance from the galaxy. As evident from Fig.~\ref{fig:L-D}, this classical FR~I/FR~II dichotomy in terms of jet power has now virtually disappeared and jets with different powers can be represented by both FR~I and FR~II morphologies. The analysis presented in \citet{Mingo} supports a widely discussed connection between the jet morphology and its inner environment (see e.g., \citealt{Gopal}). The entrainment of thermal material (e.g., ISM created in stellar winds, dense molecular clouds and atomic gases etc.) in the vicinity of jets can effectively slow down jets (see e.g., \citealt{Laing}) and may also be responsible for magnetic field evolution and particle (re)acceleration (e.g., {\citealt{Young}). This entrainment process will be less effective in high power jets where a laminar flow can be maintained over large distances. In order to understand 350-kpc  well collimated FR~I jets in NGC 5322, some important properties of the galaxy and its environment are summarized below. Although FR~I and FR~II nomenclature in the present context may not be important, we still emphasized 'FR~I' jets in NGC 5322 since neither FR~I nor FR~II radio sources were previously detected in a region in the luminosity-size diagram where NGC 5322 is found to be located. 

NGC 5322 is the foremost member of a galaxy group with velocity dispersion  of 169 km s$^{-1}$ with 21 members \citep{Makarov}.  
In a study of several galaxy groups using {\it XMM-Newton}, \citet{Fino} found that NGC 5322 is a X-ray faint (log L$_\mathrm{X}=40.41\pm0.38$ erg s$^{-1}$; \citealt{Mul}) group with exceptionally low thermal pressure in the X-ray emitting gas in the inner region. As the X-ray emission was detected only in close proximity of the central BH, it was suggested to be coming from the ISM within the galaxy and not with any in-falling matter on cluster-wide scales associated with cosmological structure formation. Therefore, the IGM surrounding NGC 5322 is likely to be of very low density. It is also worth to mention here that we found NGC 5322 to be well located in the fundamental plane of BH activity, described by a relation in \citet{Merloni} between X-ray luminosity, BH mass, and 5~GHz radio luminosity, within its scatter, using X-ray luminosity (log L$_{\mathrm X}$/erg s$^{-1}$=$40.41\pm0.38$; \citealt{Mul}), BH mass (log M$_{\mathrm{BH}}$/M$_{\odot}$=$8.41\pm0.40$ ; \citealt{Dullo}), and 5 GHz luminosity (log L$_{5\mathrm{GHz}}$/erg s$^{-1}$=$38.3\pm0.08$) using the 4.8 GHz flux as $47\pm7$ mJy from \citet{Gregory} for NGC 5322. Therefore, the BH activities in NGC 5322 appear similar to other radio galaxies in terms of central engine properties and the jet power. 

NGC 5322 is also a core-depleted galaxy with a large deficit of stellar mass in the core \citep{Dullo}. Such cores can result from dry mergers of galaxies. NGC 5322 lacks rich ISM as evidenced from non-detection of both molecular hydrogen \citep{Young} and neutral hydrogen emission \citep{Hib, Serra}. Only narrow neutral hydrogen absorption has been detected \citep{Serra}. This absorption is likely to be originating in the edge-on dusty kpc-scale disk detected in the {\it HST} image. 

With the inferred properties of ISM and IGM as above, we can speculate the following astrophysical situation for the jets in NGC 5322. The jets in NGC 5322 are undergoing weak entrainment from the ISM within the galaxy and further even much weaker entrainment outside the galaxy in the IGM. The slow-down process of the jets on parsec-scale could also be very weak due to large deficit of stellar material in the inner environment of the jets. The likely scenario for the large collimated jets in NGC 5322 therefore appears to be lack of dense material (stars or gases) near the core and low density of ISM and IGM surrounding the jets throughout its progression. This result strengthens the views presented in earlier studies (e.g., \citealt{Gopal} and \citealt{Mingo}) where the FR dichotomy is explained based upon jet interaction with the medium external to the central engine. Under an assumed theoretical scenario where the brightness evolution (re-acceleration) and the collimation of jets on kpc-scales are governed entirely by entrainment of material, the jets may remain well collimated over long distances due to lack of strong external instabilities and also become faint due to lack of particle (re)acceleration. 

\section{Summary} 

We reported here the detection of large-size ($\sim350$ kpc) faint (average surface brightness $\sim76~\mu$Jy arcsec$^{-2}$) radio jets in the nearby FR~I (total luminosity $\sim 3.7\times10^{22}$ W Hz$^{-1}$ at 144 MHz) elliptical galaxy NGC 5322 in the 144 MHz LoTSS image. Previously, radio jets were detected only up to $\sim20$ kpc extent in this galaxy. This detection highlights unprecedented sensitivity achieved by LOFAR in terms of detectable angular size at very low surface brightness. NGC 5322 is located in the radio luminosity-size diagrams in a lower-right region where previously no radio galaxy of similar size has been reported within more than one order of magnitude luminosity. The low-density environment in terms of stellar core depletion, lack of rich ISM in the galaxy and low density of IGM surrounding the jets could be identified as the main cause for long collimated jets despite of low radio luminosity in NGC 5322. This result strengthens previous works where environments around jets were found to be deciding factors for the observed radio source morphologies.  

\section*{Acknowledgements} 
LOFAR data products were provided by the LOFAR Surveys Key Science project (LSKSP; https://lofar-surveys.org/) and were derived from observations with the International LOFAR Telescope (ILT). The efforts of the LSKSP have benefited from funding from the European Research Council, NOVA, NWO, CNRS-INSU, the SURF Co-operative, the UK Science and Technology Funding Council and the Jülich Supercomputing Centre. 

This research has made use of various on-line tools and data archives - NASA Extragalactic Database, Skyview (https://skyview.gsfc.nasa.gov/), Canadian Initiative for Radio Astronomy Data Analysis (CIRADA; http://cutouts.cirada.ca), cosmology calculators (https://cosmocalc.icrar.org), FIRST and NVSS database from National Radio Astronomy Observatory, USA, SDSS database provided by the Alfred P. Sloan Foundation, the U.S. Department of Energy Office of Science, and the Participating Institutions, and the GMRT data provided by National Centre for Radio Astrophysics of the Tata Institute of Fundamental Research.   

\section*{data availability} 

All the data underlying this article are available in the article. The data used to make images can be accessed in the FITS format from the open web-based archives of various radio and optical surveys enumerated in the paper.

\bsp	
\label{lastpage}
\end{document}